\documentclass[prd,twocolumn,superscriptaddress,showpacs]{revtex4}

\usepackage{graphicx}
\usepackage{latexsym}
\usepackage{bm}
\usepackage{amsmath}

\newcommand{\comment}[1]{}

\newcommand{\Tr}{\text{Tr}}
\newcommand{\ReTr}{\text{ReTr}}

\begin{document}


\title{Gluon production in the Color Glass Condensate model of 
collisions of ultrarelativistic finite nuclei}

\author{Alex Krasnitz}
\affiliation{FCT and CENTRA, Universidade do Algarve, Campus de Gambelas,
   P-8000 Faro, Portugal}
\author{Yasushi Nara}
\affiliation{%
 RIKEN BNL Research Center, Brookhaven National Laboratory,
                Upton, N.Y. 11973, U.S.A.
}
\author{Raju Venugopalan}
\affiliation{%
 c\ Physics Department, Brookhaven National Laboratory, Upton, N.Y. 11973,
U.S.A.
}
\affiliation{%
 RIKEN BNL Research Center, Brookhaven National Laboratory,
                Upton, N.Y. 11973, U.S.A.
}

\date{\today}

\begin{abstract}
We extend previous work on high energy nuclear collisions
in the Color Glass Condensate model 
to study collisions of finite ultrarelativistic nuclei. 
The changes implemented include 
 a) imposition of color neutrality at the nucleon level
 and b) realistic nuclear matter distributions of finite nuclei.
 The saturation scale characterizing the fields of color charge is
 explicitly position dependent, $\Lambda_s=\Lambda_s(x_T)$. 
We compute gluon distributions both before and after the collisions.
The gluon distribution in the nuclear wavefunction before the collision is
  significantly suppressed below the saturation scale
 when compared to the simple McLerran-Venugopalan model prediction,
 while the behavior at large momentum $p_T\gg \Lambda_s$ remains unchanged.
We study the centrality dependence of produced gluons and compare it
 to the centrality dependence of charged hadrons exhibited by the RHIC data.
We demonstrate the geometrical scaling property of the initial gluon
 transverse momentum distributions for different centralities.
Classical Yang-Mills results for $p_T < \Lambda_s$ are simply matched to
perturbative QCD computations for $p_T > \Lambda_s$-the resulting energy
 per particle is significantly lower than the purely classical estimates.
Our results for nuclear collisions can be used as initial conditions
 for quantitative studies of the further evolution
 and possible equilibration of hot and dense gluonic matter
 produced in heavy ion collisions.
Finally, we study $pA$ collisions within the classical framework.
Our results agree well with previously derived analytical results in the appropriate kinematical regions.

\end{abstract}


\maketitle

\section{Introduction}

An outstanding problem of considerable theoretical and experimental interest
 is the possible formation of a quark gluon plasma (QGP)
 in collisions of nuclei at very high energies.
 Since the collisions are very violent
 and the time scales involved are ephemeral,
 the answer to the question of whether a QGP is formed depends sensitively on
the initial conditions for the matter produced in the collision.
Specifically, it requires an understanding of the distributions of
partons in the wavefunctions of the two nuclei {\it before} the collision.
What do these parton distributions look like? 

It is well known that multi-particle production in high energy
collisions is dominated by modes in the nuclear wavefunction
which carry a small momentum fraction $x$ of the nuclear momentum~\cite{GLRMQ}.
Understanding the correct initial conditions therefore requires
 that we understand the properties of these small $x$ modes.
 In recent years, a sophisticated effective
field theory approach has been developed to describe properties of
partons at small $x$~\cite{MV,JKMW,JIMWLK}-these form a Color Glass
Condensate (CGC)~\cite{MV,CGC,RajGavai}. The CGC is characterized by a bulk
scale $\Lambda_s$ which grows with energy and the size of the
nuclei. For RHIC energies, if one assumes $\Lambda_s$ is a constant as
for cylindrical nuclei, then $\Lambda_s\sim 1$-$2$
GeV~\cite{Mueller1,ActPol,KN,AYR02}.  The field strengths in the
saturation region behave as $\sim 1/\alpha_S$: since
$\alpha_S(\Lambda_s)\ll 1$, the field strengths are
large. Furthermore, the occupation number of saturated gluons is also
$\sim 1/\alpha_S \gg 1$. 

The initial conditions for nuclear collisions can be formulated in the 
CGC~\cite{KMW,DYEA,GyulassyMcLerran,Balitsky}.
Since the occupation numbers of the partons before the collision are large,
the very initial stage of the nuclear collision can be treated classically.
In practice, this means that one solves the Yang-Mills equations
for two sources of color charge with initial conditions given by the
classical fields of the two nuclei before the collision.
This program is difficult to accomplish analytically,
but it can be formulated and solved numerically~\cite{AR99,AR00,AR01,AYR01}.
An important assumption is that of boost invariance:
the gauge fields are assumed not to depend on the space-time rapidity $\eta$. 
The QCD Hamiltonian (in $A^\tau=0$ gauge,
 where $\tau$ is the proper time) can then be formulated as a dimensionally
 reduced 2+1-D theory of transverse gauge fields coupled to an adjoint scalar
field. Numerically, the lattice analog of the dimensionally
 reduced Hamiltonian is the Kogut--Susskind Hamiltonian~\cite{Kogut}
coupled to an adjoint scalar field~\cite{AR99}.
The space-time evolution of the classical gauge fields after the
collision is obtained by solving,
 for each configuration of the color charge sources, 
Hamilton's equations for the canonically conjugate momenta and fields.
 At late times $\tau > 1/\Lambda_s$  the system is well approximated
 as a collection of weakly coupled harmonic oscillators, and the energy and 
number distributions can be computed accordingly
 for an SU(2) gauge theory~\cite{AR00,AR01}
 and an SU(3) gauge theory~\cite{AYR01}
 after averaging over all color configurations.

In these earlier studies, 
nuclear collisions, for simplicity, were idealized as central collisions of 
infinite, cylindrical nuclei. 
The color charge squared $\Lambda_s^2$ 
was taken to be a constant for the uniform cylindrical nuclei. Furthermore, 
color neutrality was imposed only in 
a global sense~\cite{RajGavai}, namely, the color charge distribution over the entire nucleus was constrained to be 
zero. While very useful in obtaining first estimates of the space-time 
evolution of the produced gluonic matter, 
these studies did not make predictions for realistic nuclear collisions. 
In addition, studies of the distributions 
in peripheral collisions, in particular of the azimuthal anisotropy associated with elliptic flow, require 
finite nuclei 
and realistic nuclear matter distributions within each nucleus. In a recent 
paper we 
implemented all of these realistic requirements to compute elliptic flow 
in the CGC model~\cite{AYR02}. In this 
paper we will discuss the effects of imposing color neutrality  
in more detail and compute 
energy and number distributions as a function of centrality.

Strictly speaking, the results of our numerical simulations are only valid for very early 
times ($\tau\sim 1/\Lambda_s$) when the occupation numbers of the fields are large enough for 
the classical field approximation to hold. The subsequent evolution of the system can be treated 
by solving semi-classical transport equations. An interesting scenario of how thermalization may 
be achieved in this picture was discussed by Baier et al.~\cite{BMSS,BMSS2}. The results presented here can 
be used as initial conditions for detailed simulations of the equilibration dynamics in the early stage 
of heavy ion collisions~\cite{MBVDG,Bass,Nara:2001zu,ShiMuller}.

The results here can also be compared to the data on global
observables from RHIC, if one assumes parton-hadron
duality~\cite{KN,KNL,Juergen}. Remarkably, the CGC results on the
centrality and energy dependence of charged particle multiplicities
and on the rapidity and $p_T$ distributions appear to agree with
data. However, there are data on the transverse energy per particle
and on the elliptic flow ($v_2$ and $v_2(p_T)$) which appear to
require additional assumptions beyond simple parton-hadron duality. We
will discuss constraints from the data and from our simulations on our
scenarios of equilibration and hydrodynamic flow in heavy ion
collisions.

Finally, we will discuss $pA$ collisions at RHIC energies.
 This topic has recently attracted much theoretical interest
 recently~\cite{DJLT} because of the likelihood that such
 (more specifically $dA$) 
collisions will be performed at RHIC in the immediate future.
 This problem was first discussed in the classical framework by 
Kovchegov and Mueller~\cite{KovMuell}
 and subsequently by Dumitru and McLerran~\cite{Dumitru}.
 Both these authors derived analytical expressions
 for gluon production in $pA$ collisons.
 In particular, Dumitru and McLerran showed that the gluon distribution,
 while proportional to $1/p_T^4$ for $\Lambda_{s1}<\Lambda_{s2}<p_T$, is 
proportional to $1/p_T^2$
 in the kinematic region $\Lambda_{s1}<p_T<\Lambda_{s2}$.
 Here $\Lambda_{s1}$ and $\Lambda_{s2}$ are the color charges squared of
 the proton and nucleus respectively.
 Reproducing these results is a test of our numerical framework
 and we will show that our results indeed agree well with
 the analytical expressions.

This paper is organized as follows.
 In the following section,
 we will discuss the numerical simulations of nuclear 
collisions with emphasis on improvements over previous work.
 In particular, we will discuss how the condition of color neutrality is
imposed on the color charge configurations to ensure that
 there are no gluon fields and gluon production outside the nucleus.
In section~\ref{sec:mult},
 we will discuss the results of our simulations
 for gluon distributions both before and after the collision.
We compute the centrality dependence of the distribution of produced gluons
 and compare these to the RHIC data.
In section~\ref{sec:pA},
 we will discuss the results of our simulation of a $pA$ collision. 
In the final section, 
we conclude with a summary and a discussion of further improvements in our 
approach.

\section{Model of Collisions of Finite Ultrarelativistic Nuclei}
 \label{sec:model}

We will discuss in this section the McLerran-Venugopalan model (MV)~\cite{MV} for ultrarelativistic nuclei with 
realistic nuclear matter distributions. We will compute the classical gluon distributions of the nuclei both before 
and after the collision.

We first discuss parton distributions in a single nucleus in the MV
approach~\cite{MV,JKMW}.  In the original MV approach, the color
charge squared per unit area, $\Lambda_s^2$, is taken to be much
larger than the confinement scale (~$\sim \Lambda_{QCD}^2$). For
realistic nuclei at finite energies, confining effects cannot be
ignored. For instance if in a finite nucleus no color neutrality
condition is imposed, the gluon distribution outside the nucleus would
be finite even if the distribution of color charges were localized.
The issue of color neutrality in the MV model of a single nucleus was
first discussed at length by Lam and Mahlon~\cite{LM}. In their work,
they considered large nuclei with uniform nuclear matter
distributions.  In this section, we will construct color charge
distributions by imposing different color neutrality constraints at
the nucleon level.

Once we have the color charge distributions, we are ready to consider
the case of nuclear collisions. The problem was first formulated for very 
large nuclei by
Kovner, McLerran and Weigert~\cite{KMW}. The full numerical solution
of the problem for uniform cylindrical nuclei was discussed by us in a
series of papers~\cite{AR99,AR00,AR01,AYR01}.  We will discuss in this
section numerical solutions for the case of finite nuclei. Color 
neutrality constraints prove essential to avoiding particle production outside 
interaction region.

\subsection{McLerran Venugopalan model for a finite ultrarelativistic nucleus}

We will begin with a very brief review of the McLerran-Venugopalan
 model~\cite{MV,JKMW}. 
At high energies, multi-particle production is dominated by wee partons 
which carry a small fraction $x$ of the longitudinal light cone momentum
 $P^+$ of the nucleus. 
The McLerran-Venugopalan 
model is a model of the distribution of these wee partons coupled to hard parton sources 
at large $x$. In the infinite momentum frame, the light cone current of these sources 
can be expressed as  (for a nucleus moving along the positive $z$-axis) $J\equiv(J^+,0,0,0)$, 
where $J^{+,a}=\rho^a(x_T)\delta(x^-)$ is the color charge 
density on the light cone. Note that light cone co-ordinates are defined here as 
$x^{\pm} = (t\pm z)/\sqrt{2}$. It is assumed here the color charge distribution in $x^-$, due to the Lorentz 
contraction of the light cone charges, is well approximated by a delta 
function. Also, $\rho^a(x_T)$ is the  color charge distribution of the hard static 
sources (at large Bjorken $x$) in the transverse plane.
These sources are coupled to the ``soft" (small Bjorken $x$) partons and different 
color charge configurations are assumed to be distributed with the Gaussian weight  
\begin{equation}
P[\rho]=\exp\left(-\int dx^-d^2 x_T
     { \rho^a (x^-,x_T)\rho^a(x^-,x_T) \over 2\Lambda_s^2(x^-,x_T)}
  \right) \, .
\label{eqn2}
\end{equation} 
In the original McLerran-Venugopalan model, the correlator 
of color charges in the transverse plane of the nucleus satisfied the relation
\begin{equation}
\langle \rho^a (x_T) \rho^b (y_T)\rangle = \Lambda_s^2 \delta^{(2)}(x_T - y_T) \delta^{ab} \, ,
\label{eq:corr}
\end{equation}
where $\Lambda_s^2$ is a constant. As pointed out by Lam and 
Mahlon~\cite{LM}, the $\delta$-function form of the uncorrelated color charges is inconsistent with color neutrality 
at large transverse separations.
In general, we will have 
$\langle \rho^a (x^-,x_T)\rho^b (0)\rangle = \Lambda_s^2(x^-,x_T) \delta^{ab}$.

In the limit of an ultrarelativistic source, the classical 
equations of motion for the wee parton fields are
\begin{equation}
D_{\mu}F^{\mu\nu, a} = \delta^{\nu +}\rho^a (x^-,x_{T})\, , 
\end{equation}
where $F^{\mu\nu,a}=\partial_\mu A^{\nu,a} -\partial_\nu A^{\mu,a}+f^{abc}A^{\mu,b}A^{\nu,c}$ 
is the field strength tensor and $D_{\mu}=\partial_\mu + iA_\mu$ is the 
covariant derivative associated with the wee parton gauge field $A^\mu$. 
We first find a solution which satisfies
$A_{C}^i=A_C^-=0$ in the covariant gauge $\partial_{\mu}A_C^{\mu}=0$.
For this solution, the covariant gauge condition becomes $\partial_+A_C^+=0$.
Then $A_C^+$ satisfies the Poisson's equation
\begin{equation}
 -\nabla^2_{T} {A}_{C}^+ = \rho(x^-,x_T) \, .
\label{eq:poisson}
\end{equation}
Note that one can  alternatively obtain Eq.~(\ref{eq:poisson}) in $A^-=0$ gauge
with the static condition $\partial_{+}A^{\mu}=0$.
Gauge transforming our solution from covariant gauge to the light-cone gauge $A^+=0$, 
one finds the solution~\cite{MV,JKMW,LM,JIMWLK}:
\begin{equation}
 A^-=0, \quad  A^i = i U\partial^i U^{\dagger},
\quad i\partial_- U^{\dagger} = \Lambda U^{\dagger}\, ,
\label{eq:solution}
\end{equation}
where $U$ is a gauge transformation matrix in the fundamental representation. 
Here we define $\Lambda = A_C^+$. Also, note that in the light-cone gauge, $A^i$ ($i=x,y$) is a pure gauge 
in the transverse direction: $F^{xy}=0$. The solution of the last of the equations in Eq.~(\ref{eq:solution}) 
gives 
\begin{equation}
 U = {\rm P}\exp\left[ i \int^{x^-}_{-\infty} \Lambda(z^-,x_T)\, dz^-\right]\, .
\end{equation}
From Eq.~(\ref{eq:poisson}), one then has finally the solution 
\begin{eqnarray}
 A^i&=& i \left({\rm P} 
\exp\left[i \int^{x^-}_{-\infty}
  {-1\over \nabla^2}\rho(z^-,x_T) dz^-\right]\right)\nonumber \\
 &\times& \nabla^i \left( {\rm P}\exp\left[ i \int^{x^-}_{-\infty}
{-1\over \nabla^2}\rho(z^-,x_T)dz^-\right]\right)^\dagger\, . 
\end{eqnarray}
The only non-zero component of the field strength is $F^{+i}$.

\begin{figure}
\includegraphics[width=3.5in]{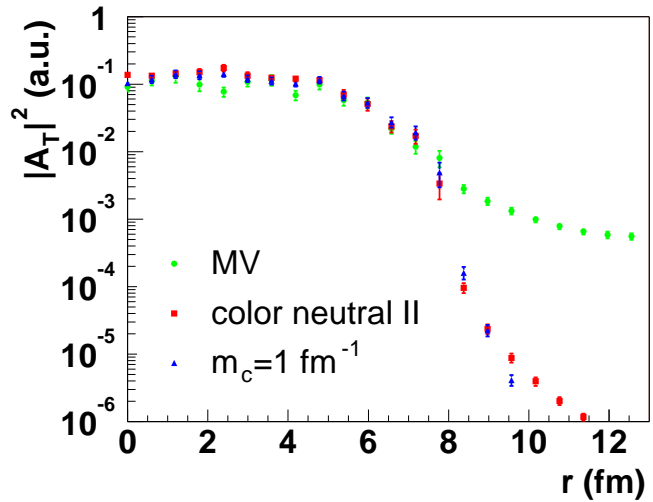}
\caption{Gluon field as a function of radial distance.
Original MV model is shown by circles,
while squares correspond to the Color Neutral II prescription (see text).
Results from Eq.~(\ref{eq:proca})  are shown by triangles.
The results are for  $\Lambda_{s0}R = 18.5$.
}
\label{fig:magfield}
\end{figure}

Thus far we have not imposed any restriction on the color charge distribution.
If we impose the simple and obvious constraint that the color charge distribution must 
be zero outside the nucleus,
 the solution of Poisson's equation in Eq.~(\ref{eq:poisson})
 can still give a non-zero gluon distributions outside the nucleus.
In two dimensions, the fall-off of the gluon field is rather slow as shown 
in Fig.~\ref{fig:magfield}.
This slow fall-off is a problem for a finite nucleus since the gluon field is 
associated with a non-zero field strength. Clearly the simple prescription 
for color neutrality 
is not sufficiently stringent.

A more realistic prescription would be to apply the color neutrality
constraint already at the nucleon level.  Our numerical procedure to implement
the constraint for finite nuclei is as follows.  We first sample $A$ nucleons
on a discrete lattice requiring that they satisfy a Woods-Saxon nuclear density profile in the transverse plane.  Note
that this procedure generates the same distribution in the continuum as \begin{equation}
\Lambda_s^2 (\bm{x}_T) = \Lambda_{s0}^2 T_A(\bm{x}_T)\, ,
\label{eqn3} \end{equation} where $T_A(\bm{x}_T) = \int^{\infty}_{-\infty} dz\,
\kappa(\bm{r})$ is a thickness function, $\bm{x}_T$ is the transverse coordinate
vector (the reference frame here being the center of the nucleus), $\kappa(\bm{r})$ is the Woods-Saxon nuclear
density profile, and $\Lambda_{s0}^2$ is the color charge squared per unit
area in the center of each nucleus. The only external dimensional variables in
the model are $\Lambda_{s0}$ and the nuclear radius $R$.

Next, Gaussian color charge distributions are generated on the lattice, where the probability 
distribution of color charge in a nucleon is expressed as 
\begin{equation}
P[\rho] = \exp\left(-\sum_j^N { \rho_j^{2}
   \over  2\Lambda_{n,j}^2} \right) \, ,
\label{eqn2}
\end{equation} 
where $\Lambda_{n,j}^2$ is the color charge distribution squared, per unit area, of a nucleon at a 
lattice site $j$ and $N$ is the number of lattice sites that comprise a nucleon. 
$\Lambda_{n,j}^2$ is obtained from $\Lambda_{s0}^2$ by assuming that 
the color charges of the nucleons add incoherently.
There are two versions of the subsequent step. In the first (which we term 
Color Neutral I), we subtract from every $\rho_j$ the spatial average
$\sum_j \rho_j/N$ in order to guarantee color neutrality 
$\langle \rho \rangle = 0$ for each 
nucleon.
In the second (termed 
Color Neutral II), the dipole moment $\bf d$
of each nucleon is eliminated in a similar manner, by superimposing the result
of the Color Neutral I procedure with a uniform distribution of dipole moments,
whose intensity, integrated over the nucleon, is $-\bf d$. This in turn
amounts to placing the appropriate charges along the line boundary of the
transverse cross-section of the nucleon.
There is {\it a priori} no reason to prefer one prescription over the 
other as long as neither leads to significant field strengths outside the 
nucleus. 

\begin{figure} \includegraphics[width=3.5in]{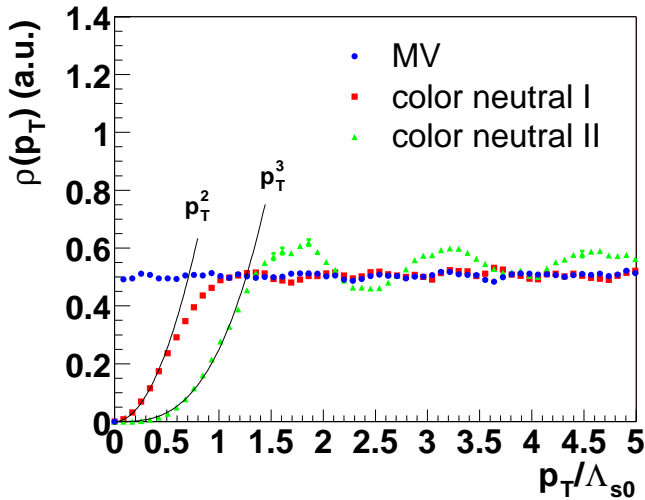}
\caption{Color Charge Correlator in momentum space.  Original MV model is
shown by circles, while squares correspond to the result from color neutrality
condition I and triangles correspond to color neutrality condition II (see
text).  The color charge correlator is plotted versus $p_T$ in units of
$\Lambda_{s0}$.  } \label{fig:colcorr} \end{figure}

In Fig.~\ref{fig:colcorr}, we plot the Fourier transform of 
the charge correlator, which in the continuum is defined as 
\begin{equation}
\tilde \rho(p_T) = \,\int d^2 x_T \exp(i\bm{p}_T\cdot\bm{x}_T)
 \langle \rho^a(x_T) \rho^a(0)\rangle \, ,
\end{equation}
for the MV model and for the two variants
 which impose color neutrality on the nucleon level. 
In the MV model,
 it is evident from Eq.~(\ref{eq:corr}) that
 $\tilde \rho (p_T)$ is a constant everywhere
 except at $p_T=0$ where it is constrained to be zero
 from the global charge constraint.
In the Color Neutral I (II) variant,
 we see that $\tilde \rho (p_T)\sim p_T^2$ ($\sim p_T^3$) 
for small momenta $p_T < \Lambda_{s0}$ and is constant at larger momenta.
The oscillatory behavior seen for 
Color Neutral I and II
is due to the fact that
the correlator in coordinate space is not strictly a delta-function.
In the coordinate space the charge correlator for the two models,
Color Neutral I and II, 
falls off rapidly,
as $\sim 1/x_T^4$ and $\sim 1/x_T^5$ respectively, at larger distances.

It is an interesting coincidence that
 the behavior of $\tilde \rho(p_T)$ in our model is similar to 
the behavior expected from the renormalization group (RG) evolution of
 color charges in the McLerran-Venugopalan model.
In Ref.~\cite{ILM}, it is shown that the screening of color charges
due to the RG evolution gives a behavior
$\tilde \rho(p_T)\sim p_T^2$ for $p_T\leq \Lambda_{s0}$
(and $\tilde\rho(p_T)$=constant for $p_T >\Lambda_{s0}$).
A recent analysis by Mueller gives the same result
with a specific prediction for the prefactor~\cite{Mueller2002}.
These results lie between the range spanned by our
``Color Neutral I'' and ``Color Neutral II''.
Since the only information on the RG in our formalism
for nuclear collisions comes from the initial conditions,
our results may be {\it quantitatively} similar to RG evolved predictions
for nuclear collisions.

We now discuss how the gauge fields are determined
 from the color charge distributions. 
The pure gauges at the site $j$ are defined with Hermitian traceless
matrices $\Lambda_j$ on the lattice as
\begin{equation}
  U_{\hat{n},j} = \exp[i\Lambda_{j+\hat{n}}]
                        \exp[-i\Lambda_j].
\label{eq:ulambda}
\end{equation}
$\Lambda_j$ can be obtained by solving the lattice Poisson equation
\begin{equation}
 -\Delta_L \Lambda_j \equiv -\sum_{n=\hat{x},\hat{y}}
  \left(
    \Lambda_{j+n} + \Lambda_{j-n}-2\Lambda_{j}
   \right)
 = \rho_j.
\label{eq:poissonL}
\end{equation}
We solve Eq.~(\ref{eq:poissonL}) numerically,
 using the fast Fourier transform (FFT) method
 which is faster than the approach employed in Ref.~\cite{AR99}.
Imposing color neutrality on the nucleon level
 (for both Color Neutral I and Color Neutral II) 
significantly suppresses the gluon field
 outside the color charge density as shown in Fig~\ref{fig:magfield}.

One can also simulate our color neutrality prescription by introducing a
screening mass $m_c$ to the Poisson equation Eq.~(\ref{eq:poissonL})
\begin{equation}
 (-\Delta_{L} + m_c^2)\Lambda_j = \rho_j, \label{eq:proca}
\end{equation}
 as a simple method to suppress the gluon field.  The range of
the screening mass must be $m_c \sim \Lambda_{QCD}$.
Although this procedure does not originate in any color-neutrality condition,
it is clear that the gluon distribution at high momentum range should not be
modified, since $|A_T(p_T)|^2 \sim 1/(p^2_T + m_c^2)$.
  Fig~\ref{fig:magfield} demonstrates that
 the gluon field distribution with $m_c=1.0 \text{ fm}^{-1}$
is nearly identical to the color neutral distribution.

\begin{figure}
\includegraphics[width=3.5in]{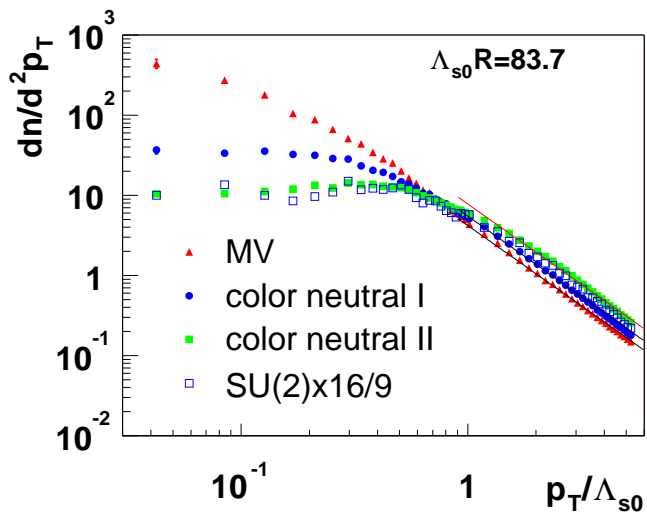}
\caption{Comparison of the unintegrated gluon distribution
 before the collision obtained by the simple MV model (triangles),
color neutral I (circles) and 
color neutral II (full squares).
The SU(2) result scaled by  16/9 is shown in open squares.
The line corresponds to a $1/p_T^2$ distribution.
Results are for $\Lambda_{s0}R=83.7$.
}
\label{fig:gluondist}
\end{figure}

In Fig.~\ref{fig:gluondist},
 we plot numerical results for the SU(3) unintegrated gluon transverse momentum
distributions before the collision.
The gluon distributions from models which impose strict color neutrality
 at the nucleon level are significantly suppressed
 at momenta $p_T< \Lambda_{s0}$ compared  to that of the original MV model
 which had only a global color neutrality constraint for the nucleus.
In the large momentum region,
 where we expect the color neutrality constraints to be immaterial,
 both sorts of models show the same $1/p_T^2$ dependence.
Comparing the SU(2) result for the Color Neutral II model to the SU(3) result,
 we observe that the ratio of the two results is the ratio of
 the respective Casimirs. 
The results shown in the figure are all for one value of
 the dimensionless coupling $\Lambda_{s0}R$.
The qualitative results are the same for other values of this parameter.

\subsection{Nuclear collisions}

We can now apply our model for parton distributions
 of finite ultrarelativistic nuclei to discuss classical gluon production
 in nuclear collisions.
 This problem was first formulated by Kovner, McLerran and Weigert~\cite{KMW}
 and perturbative solutions were discussed further by several
authors~\cite{DYEA,GyulassyMcLerran,Balitsky}.
 The numerical formulation of the classical problem has been 
discussed extensively by us in previous papers~\cite{AR99,AR00,AR01,AYR01}. In these papers, we considered, for 
simplicity, only the case of cylindrical nuclei with uniform nuclear matter distributions. Periodic boundary 
conditions were imposed. We will discuss here the formulation of the problem for finite nuclei and open boundary 
conditions. In the previous sub-section, we obtained the color charge and 
gluon distributions in coordinate space and momentum-space after applying color neutrality prescriptions at the 
nucleon level. These distributions then give us the necessary ingredients to discuss the space-time evolution of 
classical fields after the collision. For completeness (and continuity) we will repeat some of the arguments 
in the papers cited here.

We fix the gauge condition to be the radiation gauge $A^{\tau}=0$ or equivalently,
\begin{equation}
x^+ A^- + x^- A^+ = 0.
\end{equation}
We require that the light cone color charges on each light cone are delta function sources: 
$J^{\pm,a}=\delta(x^\mp)\rho^{\pm,a}(x_T)$-the solutions to the Yang-Mills equations are then 
explicitly boost invariant. We are interested in the physics pertaining to the mid-rapidity region and 
will consider the space-time rapidity $\eta$ to be zero. 
Our ansatz for the gauge field as a function of proper time $\tau=\sqrt{2x^+x^-}$ is
\begin{subequations}
\begin{eqnarray}
 A^i &=& \alpha^i_3(\tau,x_{T})\theta(x^-)\theta(x^+) \nonumber\\
       &+& \alpha^i_1(\tau,x_{T})\theta(x^-)\theta(-x^+) \nonumber\\
       &+& \alpha^i_2(\tau,x_{T})\theta(-x^-)\theta(x^+),\\
 A^{\pm} &=& \pm x^{\pm} \alpha(\tau,x_T)\theta(x^-)\theta(x^+).
\end{eqnarray}
\end{subequations}
The boundary conditions are determined by matching the solutions
in the space-like and time-like regions. Requiring that the gauge fields
must both be regular at $\tau=0$,
$D_{\mu i} F^{\mu i} = 0 $ and $D_{\mu +} F^{\mu +} = J^+$
for $x^-,x^+ \to 0$ gives the boundary conditions at $\tau=0$:
\begin{subequations}
\label{eq:matching}
\begin{eqnarray}
 \alpha^i_3(0,x_T) &=& \alpha^i_1(0,x_T) + \alpha^i_2(0,x_T),\\
 \alpha(0,x_T) &=& {i\over 2} [\alpha^i_1(0,x_T),\alpha^i_2(0,x_T)].
\end{eqnarray}
\end{subequations}
Note that these conditions, first formulated for infinitely large nuclei, 
are the same for finite nuclei. 
There is no ambiguity in determining the reaction zone 
since, for example, when $\alpha^i_1 = 0$, 
we have $\alpha = 0$, $\alpha^i_3 = \alpha^i_2$, a pure gauge-hence 
there is no classical gluon production outside the reaction zone.

\begin{figure}
\includegraphics[width=3.5in]{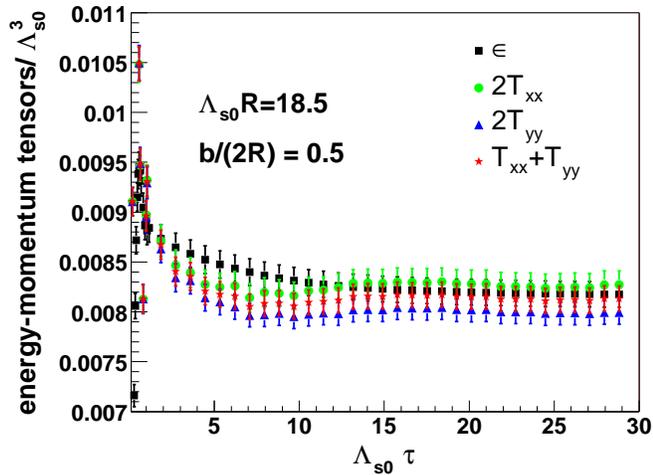}
\caption{The two components of the transverse pressure
 ($T_{xx}$ and $T_{yy}$) and the energy density 
$\epsilon$ plotted as a function of $\tau$ in dimensionless units.
 The results are for a impact parameter $b=R$
 and dimensionless coupling $\Lambda_{s0}R=18.5$.
Also shown is the sum of the two transverse pressures.}
\label{fig:stress}
\end{figure}

We shall now explain how one generates the  reaction zone in the model.
The initial conditions (at $\tau=0$) can be generated in two possible ways.
\begin{itemize}
\item[(a)]For a given impact parameter $b$, one determines the number of participant 
nucleons using the Glauber model as for example in Ref.~\cite{KN}. The corresponding 
color charge density (and the gluon distribution) for each nucleus in the reaction zone can then be determined 
using the sampling procedure described previously.
The matching condition in Eq.~(\ref{eq:matching}) is then 
used to determine the initial conditions for the evolution. 
\item[(b)]The color charge distribution is determined separately for each nucleus  by projecting 
the color charge distributions of all the color neutral nucleons in the transverse plane. The gluon distribution for 
each nucleus is then determined accordingly, and the reaction zone is automatically given by the matching condition 
in Eq.~(\ref{eq:matching}) for a specified impact parameter $b$.
\end{itemize}

The reaction zone can be determined naturally from the procedure (b)
as long as we have the correct color charge distribution for a finite nucleus,
since the region of $A_1=0$, $A_2$ remains pure gauge.
Nevertheless, we would like to use (a) for  purely technical reasons,
namely, since it is more efficient for numerical computations. We have checked that the two procedures 
give nearly identical results.

Explicit expressions for the discretized lattice Hamiltonian, for 
other components of the stress-energy tensor, and the classical equations of motion 
can be found in the appendix. In Fig.~\ref{fig:stress}, we plot the temporal behavior of the two 
transverse pressures and the energy density.
 Interestingly, at late times we notice that 
$\epsilon=T_{xx} + T_{yy}$ as one would anticipate for a free streaming gas of gluons in the 
transverse plane. We will return to this result later. 
The difference between the two transverse pressures,
 $\epsilon_p = T_{xx} -T_{yy}$ is a 
measure of the azimuthal anisotropy generated by classical fields~\cite{AYR02}.

\section{Impact parameter dependence of multiplicity distributions}
\label{sec:mult}

In previous papers~\cite{AR00,AR01,AYR01},
 we compared our results for cylindrical nuclei with periodic boundary 
conditions and uniform nuclear matter distributions to the RHIC data. 
In this section, we will apply the more realistic framework developed in the previous section to compute 
the gluon multiplicity distributions as a function of centrality. If one assumes parton-hadron 
duality, these distributions can be compared, for instance,
 to the centrality dependence of the charged hadron 
spectra in heavy ion collisions at RHIC~\cite{KN}.

At late times $\tau\gg 1/\Lambda_{s0}$, the dynamics of the
classical Yang-Mills fields produced in the collision can be 
linearized and approximated by that of
a system of weakly coupled harmonic oscillators. In previous 
papers~\cite{AR01,AYR01}, 
we discussed two methods to compute the gluon number.  Both of these are applicable when the harmonic oscillator 
approximation is a good one. One method is a gauge invariant relaxation method which can only be used to 
measure the particle number and not its distribution with momenta. The other method is to fix the Coulomb gauge 
and compute the field amplitudes squared in momentum space. We showed that both methods give results for the 
gluon number which agree very well with each other.

We will therefore restrict ourselves in this paper
 to using one method alone-that of fixing the transverse 
Coulomb gauge  $\bm{\nabla}_{\perp}\cdot \bm{A}=0$~\cite{AR01} and computing the field amplitudes. We have 
\begin{equation}
N(k) =  \omega(k) \langle |\phi(k)|^2\rangle =  \sqrt{\langle
|\phi(k)^2|\rangle\langle|\pi(k)|^2\rangle}, \label{eq:coulomb} \end{equation} where
$\phi(k)$ and $\pi(k)$ correspond to the potential term and the kinetic term
in the Hamiltonian respectively.  The average $\langle \rangle$ represents an average over the
initial conditions.

\begin{table}
\caption{$\Lambda_{s0}=1.41$ GeV.
Results are for a $256\times256$ lattice-the nuclear radius is 64 lattice units. 
All dimensionful scales are in GeV units unless otherwise stated.
}
\begin{ruledtabular}
\begin{tabular}{lllllll}
 $b(fm)$ &  $N_{part}$  & $g^2N_g$ & $g^2E_g$ &
$\Lambda(b)$ & $Q_s(b)$ & $f_N(b)$ \\\hline
0.000 & 377.89 & 1628.68 & 2725.40 & 1.1777 & 1.0836 & 0.3695 \\
3.150 & 321.35 & 1309.83 & 2170.08 & 1.1545 & 1.0563 & 0.3484 \\
4.725 & 263.33 & 1035.84 & 1663.88 & 1.1234 & 1.0197 & 0.3359 \\
6.300 & 199.11 & 760.95  & 1182.02 & 1.0741 & 0.9623 & 0.3249 \\
7.875 & 136.47 & 515.61  & 751.902 & 0.9993 & 0.8758 & 0.3152 \\
9.450 & 81.21  & 295.76  & 384.004 & 0.8876 & 0.7487 & 0.3004 
\label{table:l1}
\end{tabular}
\end{ruledtabular}
\end{table}

\begin{table} \caption{$\Lambda_{s0}=2.32$ GeV.  Results are for 
a $512\times512$ lattice-the nuclear radius is 128 in lattice units. 
All dimensionful scales are in GeV units unless otherwise stated. }
\begin{ruledtabular} \begin{tabular}{lllllll} $b(fm)$  & $N_{part}$ & $g^2N_g$ &
$g^2E_g$ & $\Lambda(b)$ & $Q_s(b)$ &  $f_N(b)$ \\\hline
0.000 & 377.89 & 3768.00 & 9198.736 & 1.9517 &2.0355 &0.2867 \\ 
3.150 & 321.35 & 3061.59 & 7492.084 & 1.9132 &1.9866 &0.2757 \\
6.300 & 199.11 & 1808.89 & 4183.888 & 1.7800 &1.8185 &0.2610 \\
7.875 & 136.47 & 1215.17 & 2636.300 & 1.6560 &1.6636 &0.2522 \\
8.367 & 118.17 & 1042.54 & 2243.692 & 1.6060 &1.6017 &0.2514 \\
9.450 & 81.21  & 699.95  & 1411.900 & 1.4708 &1.4356 &0.2411 
\label{table:l2}
\end{tabular}
\end{ruledtabular}
\end{table}

In Refs.~\cite{AR01,AYR01}, non-perturbative relations were derived for the energy 
and number of produced gluons (at central rapidities) as a function of $\Lambda_s^2$. 
For realistic nuclei, these non-perturbative relations 
are less simple. One can parametrize our results for the gluon number with 
the more general relation
\begin{equation}
{dN_g\over d\eta} = f_N(b)\int d^2x_T {\Lambda_s^2(b,x_T) \over g^2},
\end{equation}
where $\Lambda_s(b,x_T)$ is the local saturation scale defined to be
$\Lambda_s^2(b,x_T)= C\cdot {\tilde{\rho}}(b,x_T)/2$,
 where ${\tilde{\rho}}$ is the 
participant density at a particular position in the transverse plane, and $C$ is 
the color charge squared per nucleon. When 
$\Lambda_s(b,x_T)$=constant, as for cylindrical uniform nuclei, one recovers 
the form of the expressions in Refs.~\cite{AR01,AYR01}. The color charge squared
in the center of the nucleus is $\Lambda_{s0}^2 = C\cdot {\tilde{\rho}}(0,0)/2$, so 
$\Lambda_s^2(b,x_T)=\Lambda_{s0}^2 {\tilde{\rho}}(b,x_T)/{\tilde{\rho}}(0,0)$. 
One can then re-write the previous equation as 
\begin{equation}
{dN_g\over d\eta}
    = {f_N(b) \over g^2} {\Lambda_{s0}^2\over \rho_0}N_{part}(b),
\label{eq:npart}
\end{equation}
where $\rho_0 = {\tilde{\rho}}(0,0) = 4.321 \text{ fm}^{-2}$
 and $N_{part}=\int d^2 x_T {\tilde{\rho}}(b,x_T)$.

In Tables~\ref{table:l1} and \ref{table:l2}, we show 
the calculated SU(3) results for two values of the saturation scale in the 
center of the nucleus: $\Lambda_{s0}=1.41$ and
$\Lambda_{s0}=2.32$ GeV respectively.
In the tables, $b$ is an impact parameter (in units of fm) and $N_{part}$
is a number of participants at that impact parameter. The latter 
is calculated using a Woods-Saxon nuclear density profile.
We list in the tables our results, as a function of impact parameter,
for $g^2 N_g$; the number of produced gluons 
and
 $g^2 E_g$; the transverse energy of produced gluons in GeV
multiplied by the value of the strong coupling constant squared
$g^2$, evaluated (to one loop order)
at the average value of the saturation scale
(denoted in the tables as $Q_s(b)$) for that impact parameter.

In Table~\ref{table:l1}, we see that the gluon multiplicity, for the most 
central collisions, is about half the final pion multiplicity at RHIC while 
the transverse energy of gluons is close to the experimentally observed transverse 
energy of about $600$ GeV. In Table~\ref{table:l2}, the gluon multiplicity is 
close to the final pion multiplicity at RHIC while now the transverse energy 
is nearly four times larger than the experimentally observed transverse energy. 
Neither the initial transverse energy nor the initial gluon multiplicity are 
conserved quantities, so the results for different values of the saturation 
scale lend themselves to very different dynamical scenarios. For instance, the authors of 
Ref.~\cite{BMSS} have suggested that inelastic $gg\rightarrow ggg$ processes may be 
important in the early stages of nuclear collisions thereby increasing the gluon multiplicity 
prior to thermalization/hadronization. Alternatively, if the gluon number close to the 
experimental number (as in Table~\ref{table:l2}), strong hydrodynamic 
expansion at early times might be 
essential to lower the initial transverse energy to the experimentally desired value. 
Additional observables (such as $v_2$) may be important in distinguishing between the 
different scenarios.

\begin{figure}
\includegraphics[width=3.4in]{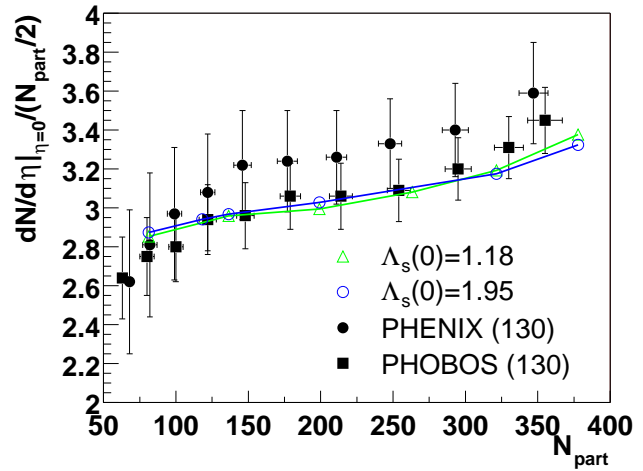}
\caption{Comparison of the centrality dependence of the gluon distribution
from SU(3) lattice results to data from
experiments~\cite{phenix_mult,phobos130}.  The strong coupling constant is
fixed to the value $g^2=4$.  The lattice results for $\Lambda_s(0)=1.18$ GeV
and $\Lambda_s(0)=1.95$ GeV
are multiplied by a factor 2.4 and 1.1, respectively.
}\label{fig:multcentdep}
\end{figure}
%
%
In Fig~\ref{fig:multcentdep},
we plot the centrality dependence of gluons calculated using
Eq.~(\ref{eq:coulomb}) together with the experimental data
from PHOBOS~\cite{phobos130} and PHENIX~\cite{phenix_mult}.
We assume here that the charged particle multiplicity is
two thirds of the gluon number. 
The classical computation is performed for fixed $\alpha_s$;
the centrality dependence, as seen from Eq.~(\ref{eq:npart}),
comes from the dependence of $f_N$ on the impact parameter.
In Ref.~\cite{AR01},
it was shown that $f_N\equiv f_N(\Lambda_s R)$ increases slowly
with $\Lambda_s R$-hence one expects it to vary with impact parameter.
We see that the results agree reasonably well with the data.

\begin{figure}
\includegraphics[width=3.4in]{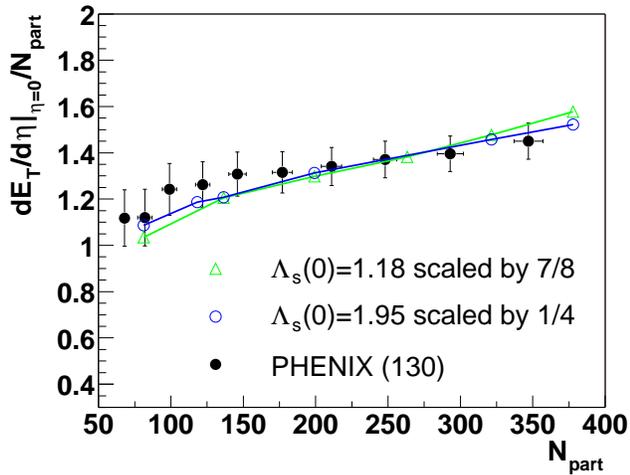}
\caption{Comparison of the centrality dependence of the
gluon transverse energy distribution from SU(3) lattice results
to data from experiments~\cite{phenix_et}.
The strong coupling constant is fixed  to the value $g^2=4$.
The lattice results for $\Lambda_{s}(0)=1.95$ GeV
are scaled by ${1\over 4}$ while those for $\Lambda_s(0) =1.18$ are scaled 
by ${7\over 8}$.
}
\label{fig:etcentdep}
\end{figure}

The centrality dependence of the transverse energy is studied in Fig.~\ref{fig:etcentdep}.
As in the case of the multiplicity, even though the absolute normalization is 
strongly dependent on one's choice of $\Lambda_s$, the centrality dependence is 
very similar for the two $\Lambda_s$'s and shows reasonable agreement with the data.

Let us now compare our results with those derived in Ref.~\cite{KNL} and discussed
further in Ref.~\cite{BMSS2}.  In these works, one obtains in terms of
${\bar{Q_s}}$, the average saturation scale, the result \begin{eqnarray}
{dN_g\over d\eta} &=& c_N \frac{N_c^2-1}{4\pi^2 N_c} \int d^2x_T {Q_s^2(b,x_T)
\over \alpha_s} \nonumber\\ &\approx& c_N xG(x,{\bar {Q}_s}^2(b))
\frac{N_{part}}{2} \, .  \end{eqnarray} In the leading logarithmic
approximation, if $c_N$ is constant, one obtains a logarithmic dependence on
the centrality entirely from $xG(x,{\bar{Q}_s^2(b)})$.  One could thus
attribute the logarithmic behavior at the classical level to fixed $\alpha_s$
and leading logarithmic behavior of the gluon distribution function or
equivalently, at higher order, to the one loop running of $\alpha_s$. In the
physically interesting regime, it is difficult to distinguish between the two.
A reasonable agreement with the data is also seen in this formulation of the
problem. 

The relation between the two formulations is as follows. 
As discussed previously, what we simulate numerically
is the color charge squared per unit area,
$\Lambda_s^2$~\cite{AR01}. The saturation scale, on the other hand, is
a scale determining the behaviour of the gluon number
distribution~\cite{KovMuell}.  The two scales can be related by
computing, for instance, the gluon number distribution in a nucleus 
in the McLerran-Venugopalan approach (where $\Lambda_s$ naturally
appears) and in Mueller's dipole approach. On comparing the two one obtains, to leading
logarithmic accuracy, the following relation between $\Lambda_s$ and $Q_s$:
\begin{equation}
   Q^2_s =  {N_c \Lambda_s^2\over 4\pi}
            \log\left({\Lambda_s^2\over \Lambda^2_{QCD}}\right),
\label{eq:qsls}
\end{equation}
with $\Lambda_{QCD}=0.2$ GeV. The relation between $c_N$ and $f_N$ is
$c_N=4\pi^2 f_N / (N_c^2-1)/ \ln(Q_s^2/\Lambda_{QCD}^2)$. Therefore,
if $c_N$ is to be a constant, $f_N$ increases logarithmically with
$Q_s$. A weak rise in $f_N$ is seen in our simulations. If the
infrared scale, as argued in Ref.~\cite{KIIM}, is a number of order
$O(1/Q_s)$, we would have $\Lambda_s\approx Q_s$.  Note that in
practice $Q_s^2(b)$ and $\Lambda_s^2(b)$ (related by
Eq.~(\ref{eq:qsls}) are nearly equal to each other for all impact
parameters.

In Ref.~\cite{BMSS}, it was argued that subsequent inelastic scattering of 
gluons after the classical approximation is applicable but 
before thermalization would increase the multiplicity by a factor 
$1/\alpha_s^{2/5}$. This effect was taken into account in Ref.~\cite{BMSS2} and 
it was shown that good fits to the data could be obtained for this functional form 
as well. The effects of parton multiplication {\it a la} Ref.~\cite{BMSS2} are beyond the 
scope of this work and we will not discuss it further here. 

\begin{figure}
\includegraphics[width=3.4in]{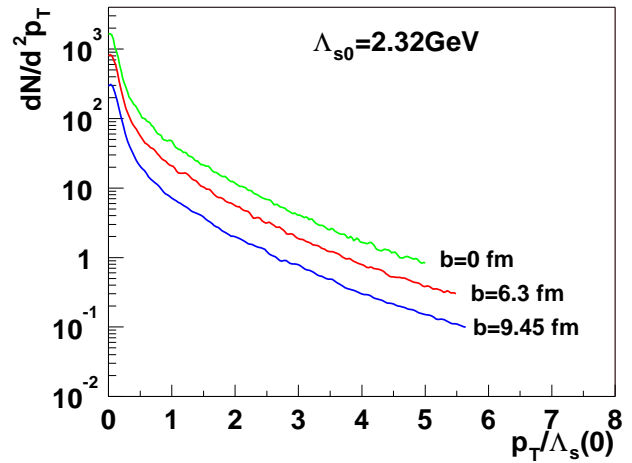}
\includegraphics[width=3.4in]{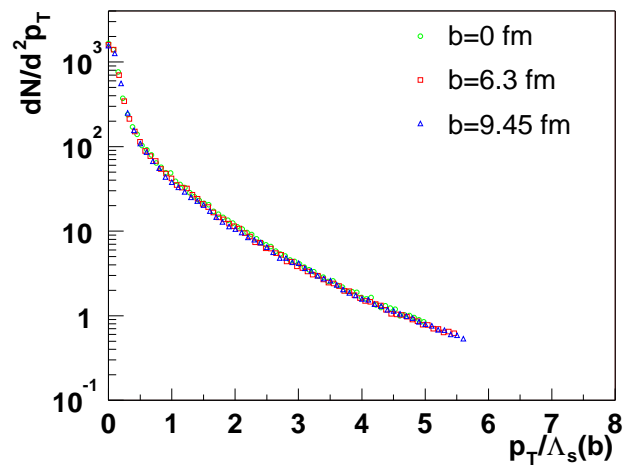}
\caption{Top: Gluon transverse momentum distributions 
as a function of $p_{T}/\Lambda_s(b)$
for $b=0.0$, 6.4, and 9.45 fm. Bottom: Same as top figure but with 
re-scaled normalizations. 
}
\label{fig:dp512}
\end{figure}

%
%

We now turn to the computation of the initial gluon transverse momentum
 distributions.
In the top half of Fig.~\ref{fig:dp512},
 we show the initial gluon transverse momentum distributions
 for different impact parameters in dimensionless units of $p_T/\Lambda_s(b)$.
In the bottom half of the figure
we plot the same quantity by re-scaling the overall normalization
 of the distributions.
 Strikingly, as a consequence of geometrical scaling,
 the distributions now coincide-the initial gluon distributions
are a universal function of $p_T/\Lambda_s(b)$.
In Ref.~\cite{Juergen}, it was argued that
the RHIC data display such a geometrical scaling.
More detailed comparisons with the recent RHIC data
at $\sqrt{s_{NN}}=200$ GeV will help further clarify the observation of 
geometrical scaling.
Note however that the typical $\langle p_T\rangle$ of the initial
gluon distribution is of order $\Lambda_s$, while 
the observed $\langle p_T\rangle$ of charged hadrons is smaller,
 of the order $3 T_c$. For the $\Lambda_s$ in Table~\ref{table:l1},
 this difference is a factor of two; for those in Table~\ref{table:l2},
it is significantly larger. 

\begin{figure}
\includegraphics[width=3.4in]{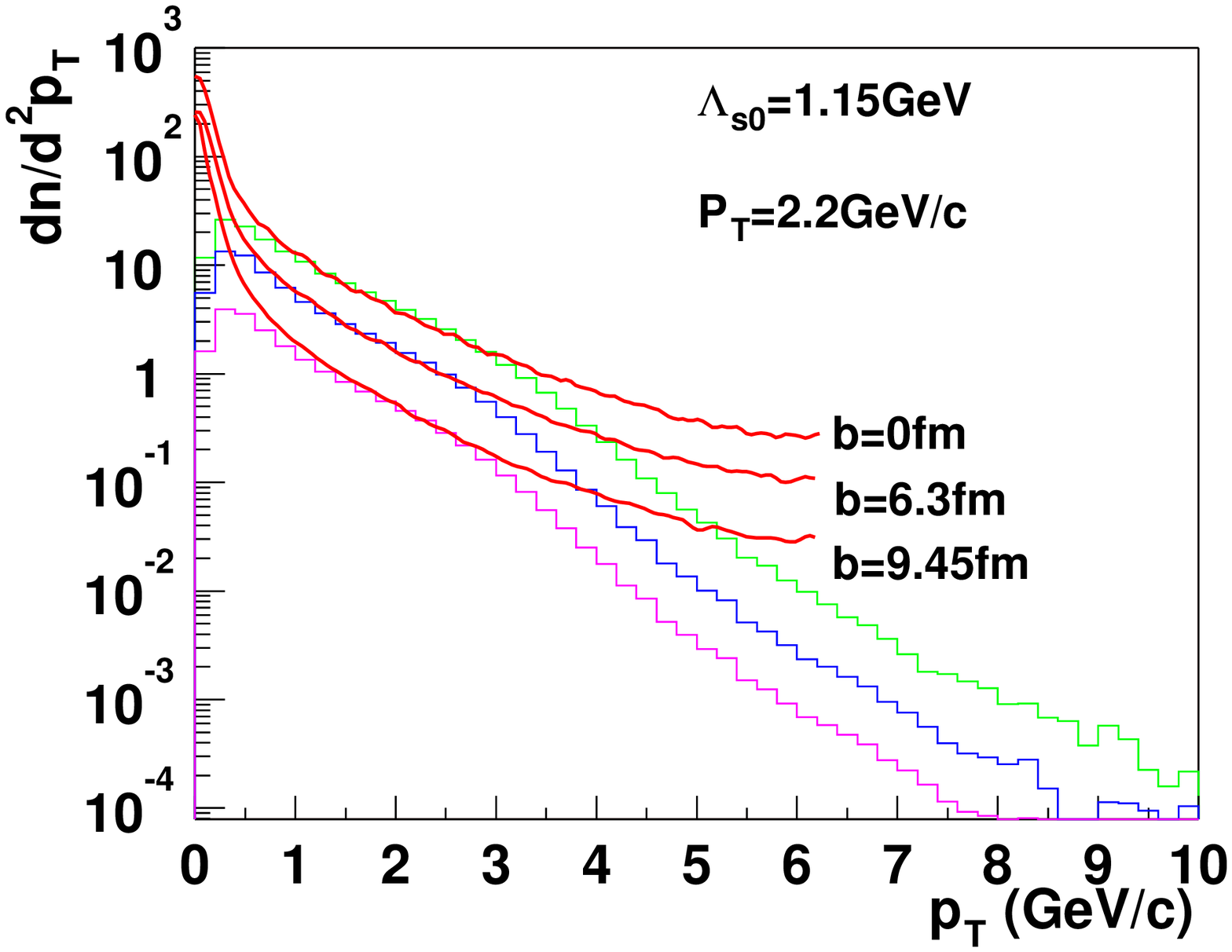}
\includegraphics[width=3.4in]{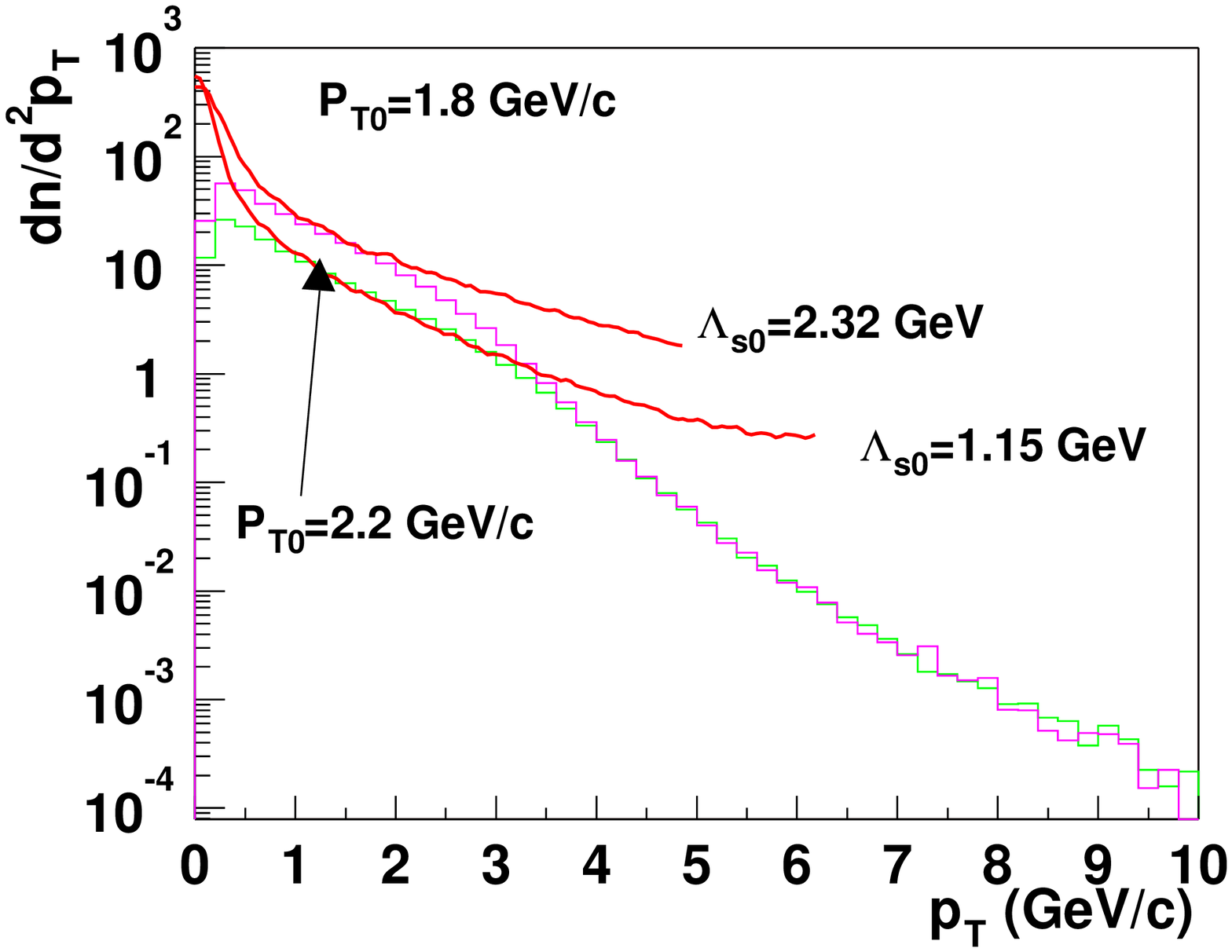}
\caption{
Top: Gluon transverse momentum distributions 
at $\eta=0$ for  impact parameters $b=0$, 6.3, 9.45 fm
in the case of $\Lambda_{s0}=1.15$ GeV.
Mini-jet partons from PYTHIA, multiplied by number of collisions, 
with cut-off momentum $p_{T}=2.2$ GeV/c, are also plotted in histograms.
Bottom: The same for impact parameters $b=0$ fm
for $\Lambda_{s0}=1.15$ GeV and $\Lambda_{s0}=2.23$ GeV.
Different cut-off momenta of $p_{T}=2.2$ and $p_{T}=1.8$ GeV/c for the 
PYTHIA mini-jets are also plotted by histograms.
}
\label{fig:ymjet1}
\end{figure}
%
%

It is instructive to compare the transverse momentum distributions
 computed in the classical approach with the pQCD computations
 of the gluon distribution.
The latter is obtained as follows.
In the parton model, when the parton density is small,
the cross section for hard parton scattering can be calculated as
 \begin{equation} {d \sigma_{jet}\over d
p^2_Tdy_1dy_2} = K\sum_{a,b} x_1x_2f_a(x_1,Q^2)f_b(x_2,Q^2)
{d\sigma_{ab} \over d\hat{t}}\, ,
\label{eq:pqcd}
 \end{equation}
 where $y_1$ and $y_2$ are the
rapidities of the scattered partons and $x_1$ and $x_2$ are the fractions of
momentum of the initial partons.  The structure functions, $f_a(x,Q^2)$ are
taken to be the CTEQ5 leading order parametrizations~\cite{cteq5}.  
The parton distributions are evaluated at the scale $Q^2= p_T^2$.
Here  $d\sigma_{ab}/d\hat{t}$ is the leading order QCD partonic cross section.
In addition, one includes initial and final state radiation processes
in order to take into account the enhancement of higher-order
contributions associated with multiple small-angle parton emission.
This is for instance what is done in the event generator PYTHIA and we use for 
our purposes PYTHIA version 6.2~\cite{pythia}
for the incident c.m. energy of $\sqrt{s}=130$ GeV.
The $K$ factor of $K=2$ is chosen to fit the UA1 data of $p\bar{p}$
at $\sqrt{s}=200$ GeV~\cite{ua1}. The nuclear 
gluon distributions are obtained by multiplying the PYTHIA result by the number of
hard binary collisions. Nuclear shadowing effects are neglected.

In the top half of Fig.~\ref{fig:ymjet1}, we study the impact parameter dependence of the
classical gluon transverse momentum distributions (with $\Lambda_{s0}=1.14$ GeV)
together with the parton distributions from PYTHIA calculations
with an infrared cut-off parameter of $p_T=2.2$ GeV/c.
We find that there is a overlap region between 1-3 GeV/c between the two computations. For higher 
transverse momenta, the classical computation overshoots the PYTHIA result. One reason this is 
the case is that the realistic parton distributions have not been included in the classical 
distribution-this makes a difference at large transverse momenta~\cite{GyulassyMcLerran,DJLT}.  
Naively, one would expect the pQCD result to diverge rapidly
at small $p_T$. The reason it doesnt is that
initial and final state radiation processes change
the parton distributions from Eq.~(\ref{eq:pqcd}) below 3-4 GeV/c.
In the lower half of Fig.~\ref{fig:ymjet1},
we plot the gluon distributions
with the parameters $\Lambda_{s0}= 2.32$ and 1.14 GeV
along with the PYTHIA gluon distributions for 
different values of the infrared cut-off.
 The latter is not an infrared safe quantity so it is not 
too surprising that the shape and magnitude of the distribution
 does depend on the cut-off. 
So if one were to match the classical and pQCD distributions,
 one has to adjust the $p_T$ cut-off for different values of $\Lambda_{s0}$.
 The differences in the gluon number
 between the classical and classical+PYTHIA LO pQCD result
 are not very significant for the particle number
 but may be so for the transverse energy since the latter is
 more ultraviolet sensitive. Therefore the large transverse energy per particle obtained in the 
classical approach may be significantly overestimated. A more realistic calculation of the 
distribution of hard modes will find less of an excess (relative to experiment) in the transverse 
energy.

We discussed in this section our results
 from numerical simulations of very high energy nuclear collisions
of nuclei at various impact parameters.
 While the absolute numbers from our results depend very sensitively 
on the scale $\Lambda_s$,
 the dependence of the gluon number and transverse energy
 on centrality and the phenomenon of geometrical scaling
 of the momentum distributions are
 independent of $\Lambda_s$ and appear universal.
In addition, they appear to agree reasonably well with trends in the RHIC data.
 Similar observations were made in Refs.~\cite{KN,KNL}
and in Ref.~\cite{Juergen} for global features of the data.

\section{proton-nucleus collisions} \label{sec:pA}

Collisions of protons/deuterium with large nuclei at RHIC energies
 are essential to distinguish dynamical effects
 due to correlations in the nuclear wavefunction
 from dynamical effects resulting 
from the collective expansion of hot and dense matter
 in nucleus-nucleus collisions.
The first computation, in the classical framework,
 of gluon production was performed by Kovchegov and Mueller~\cite{KovMuell}. 
Here we will follow the subsequent work of Dumitru and McLerran~\cite{Dumitru}
 which is closer in its formulation to the discussion here.
In Ref.~\cite{Dumitru}, the authors introduce two saturation scales 
which we will denote as $\Lambda_{s1}$ (for the proton)
 and $\Lambda_{s2}$ for the nucleus.
 One expects that since the parton density in a nucleus is
 larger than in a proton, $\Lambda_{s1}^2 \ll \Lambda_{s2}^2$. 
In the kinematic region $p_T > \Lambda_{s1}$,
 analytical computations of gluon production are feasible.
For $\Lambda_{s1}<\Lambda_{s2}<p_T$,
 the situation is analogous to nucleus-nucleus collisions
 and the distributions are proportional to
 $\Lambda_{s1}^2\Lambda_{s2}^2/p_T^4$.
 For $\Lambda_{s1}<p_T<\Lambda_{s2}$, 
the distribution changes qualitatively
 and is proportional to $\Lambda_{s1}^2/p_T^2$. In this kinematic region, 
one obtains~\cite{Ad:note} $dN/d^2 b\, d^2 k_t\propto \Lambda_{s1}^2\ln(k_t^2/\Lambda_{s1}^2)/k_t^2$.

\begin{figure}
\includegraphics[width=3.4in]{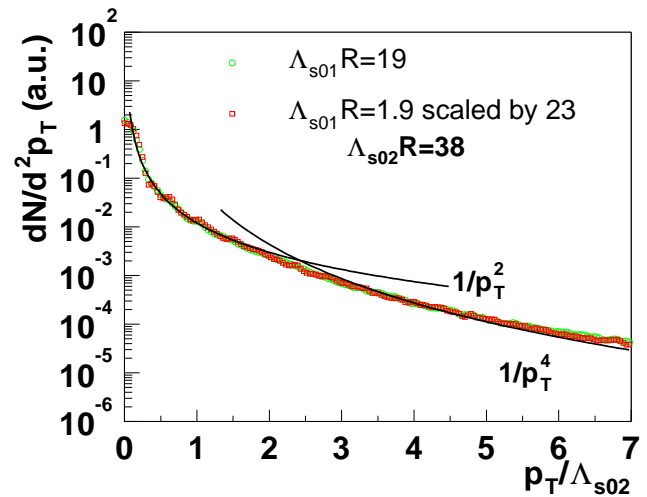}
\caption{
Transverse momentum distribution of produced gluons in a pA collision. Here $\Lambda_{s01}^2$ refers to 
the color charge squared per unit area in the center of the proton while $\Lambda_{s02}^2$ is the 
color charge squared per unit area in the center of a nucleus.
}
\label{fig:pAmatter}
\end{figure}

In our framework, $pA$ collisions can be simulated straightforwardly since the only change 
from nucleus-nucleus collisions is to consider asymmetric values of $\Lambda_s$.
The results of our simulations are shown in Fig.~\ref{fig:pAmatter}. 
We choose a fixed value for $\Lambda_{s02}$ in the center of the nucleus and plot the momentum distributions 
for different $\Lambda_{s01}$'s. At large $p_T$, as anticipated, the distribution goes as $1/p_T^4$.
At smaller $p_T$ of $p_T\sim 2\Lambda_{s02}$ one sees a qualitative change in the distributions which is 
more consistent with a $1/p_T^2$ behavior of the distributions. Thus it appears that our 
distributions reproduce the shapes predicted by 
Dumitru and McLerran in the appropriate kinematical regions. 

A comparison of absolute normalizations is more tricky. 
If one computes analytically~\cite{Ad:note} the ratio of the multiplicities at 
the scale $k_t\sim \Lambda_{s02}$ for the two values of $\Lambda_{s01}$ shown in Fig.~\ref{fig:pAmatter}, 
the ratio is $\sim 23$, which is the scaling factor used to scale the distribution with the 
lower $\Lambda_{s01}$ to that with 
the larger value. However, we should caution that the absolute normalization of these results is uncertain
 because the behavior in the kinematic region $p_T<\Lambda_{s01}$
 cannot be computed analytically~\cite{comment1} and can only be computed numerically as discussed here. 

More detailed studies of pA collisions are underway and will be reported separately~\cite{AYR03}.

\section{Summary}\label{sec:summary}

In this paper, we extended previous results for collisions of uniform, cylindrical nuclei with periodic boundary 
conditions to finite nuclei with realistic nuclear matter distributions and open boundary conditions. To 
achieve this, we imposed color neutrality constraints on nucleon color charge distributions. With these improvements, 
we were able to study nuclear collisions at different impact parameter and proton-nucleus collisions. We obtained 
results for the initial gluon number and transverse energy and compared these to the data. While the absolute 
normalizations of these quantities depend sensitively on the saturation scale $\Lambda_s$, their centrality 
dependence appears universal and shows reasonable agreement with the experimental data. Similarly, the 
transverse momentum distributions display geometrical scaling. (The latter has been argued to be a feature of 
the RHIC data~\cite{Juergen}.) Our results however also suggest that dynamical effects at later times (possibly 
beyond the validity of the classical approach) might also be important in describing the RHIC data in detail. 
This is suggested for instance by the necessity of reproducing both the transverse energy and the 
multiplicity but more emphatically by the failure of the classical approach to explain the elliptic 
flow data~\cite{AYR02}. It has been argued that the elliptic flow data may be explained by non-flow 
correlations~\cite{KT} but more detailed theoretical and experimental analyses are required to clarify 
the issue. 

We also briefly discussed proton-nucleus collisions in the classical framework. We showed that our 
numerical simulations agree qualitatively with previous analytical work~\cite{KovMuell,Dumitru}. This 
agreement suggests that results of our numerical simulations for nucleus-nucleus collisions (where 
analytical computations are more difficult) are robust.

The classical distributions discussed here can be used as the initial conditions for the
subsequent time evolution of produced gluons as might be described, for instance, in 
Boltzmann type transport calculations~\cite{Mueller1,MBVDG,BMSS,Nara:2001zu}
 at late times where field strengths are weak.
 With these initial conditions, it will be interesting to study
the consequences of parton evolution on flow or parton energy loss.

The classical treatment of nuclear collisions can also be improved significantly. One such improvement is 
to give up the assumption of strict boost invariance and to study the effects of doing so on our 
distributions. Another is to go beyond the simple assumption of Gaussian-distributed color charges 
and to incorporate the effects of renormalization group evolution of the color charges at high energies. 
All of these studies, tempered by the large amount of available experimental data, will hopefully 
provide insight into the very earliest moments of a high energy nuclear collision and of the subsequent
dynamics of the produced quark gluon matter.

\begin{acknowledgments}
We would like to thank A.~Dumitru, D.~Kharzeev, L.~McLerran,
A.~Mueller, R. Seto and X.~N.~Wang for useful comments.  R.~V.'s
research was supported by DOE Contract No. DE-AC02-98CH10886. A.K. and
R.V. acknowledge support from the Portuguese FCT, under grants
CERN/P/FIS/40108/2000 and CERN/FIS/43717/2001.  R.V. and
Y.N. acknowledge RIKEN-BNL for support. 
We also acknowledge support in part of NSF Grant No. PHY99-07949.

\end{acknowledgments}

\appendix*

\section{Space-time evolution of the Stress-Energy tensor
             in a nuclear collision}

For completeness, we provide in this appendix explicit expressions for the lattice Hamiltonian, the 
lattice equations of motion and discretized expressions for other components of 
the Stress-Energy tensor for the problem of interest. 

\subsection{Hamiltonian approach in the transverse plane}

In this sub-section, we write down the classical Hamiltonian's equations
of motion in the $\tau,\eta,\vec{r}$ coordinates where
$\tau =\sqrt{2x^+x^-}$ is the proper time, $\eta={1\over2}\log(x^+/x^-)$
is the space-time rapidity and $\vec{r} = (x,y)$ are two transverse
Euclidean coordinates. Inverse transformations are
$t=\tau \cosh\eta$ and $z=\tau \sinh\eta$ respectively.
The QCD action for gauge field in the $\tau, \eta, \vec{r}$
coordinates reads
\begin{equation}
 S_{QCD} = \int \tau d\eta d\tau d^2r \left[
            -{1\over8}\Tr(g^{\mu\alpha}g^{\nu\beta}F_{\alpha\beta}F_{\mu\nu})
        \right],
\end{equation}
where
$F_{\mu\nu}=\partial_{\mu}A_{\nu}-\partial_{\nu}A_{\mu}-ig[A_{\mu},A_{\nu}]$
and the metric is diagonal with
 $g^{\tau\tau}=-g^{xx}=-g^{yy}=1$ and $g^{\eta\eta}=-1/\tau^2$.
$A_{\mu} \equiv A_{\mu}^a t^a$, and $t^a$
represent a gauge group matrices with the normalization
of $\Tr(t^at^b) = 2\delta_{ij}$.
The Lagrangian density in $A^{\tau}=0$ gauge is 
\begin{equation}
 {\cal L}  =  \Tr\left( 
            {\tau\over4}(\partial_{\tau}A_i)^2
           +{1\over4\tau}(\partial_{\tau}A_{\eta})^2
           -{\tau\over8} F_{ij}^2 
           -{1\over 4\tau}F_{\eta i}^2 \right) \,,
\end{equation}
where $i,j$ runs over transverse coordinate $x$ and $y$.

Now let us assume $\eta$ independence of the fields. As discussed previously, the 
Yang-Mills equations have this property if the sources are strictly $\delta$-function 
sources on the light cone. We have 
\begin{equation}
  A_i(\tau,\eta,{\vec r}) = A_i(\tau,{\vec r}), \qquad
  A_{\eta}(\tau,\eta,{\vec r}) = \Phi(\tau,{\vec r}) \, ,
\end{equation}
resulting in $F_{\eta i} = -D_i\Phi$,
where $D_i = \partial_i - ig[A_i,\cdots]$ is the covariant derivative
in the adjoint representation.
Defining the conjugate momenta
\begin{equation}
  E_i =  \tau\partial_{\tau}A_i, \quad
  p_{\eta}  =  {1\over\tau}\partial_{\tau}A_{\eta}\,,
\end{equation}
one finds that the 
boost invariant Yang-Mills Hamiltonian is the QCD Hamiltonian 
in 2+1 dimensions coupled to an adjoint scalar~\cite{AR99}:
\begin{equation}
 H = \int d^2r \Tr \left\{ 
            {1\over4\tau} E_i^2
           +{\tau\over4} p_{\eta}^2
           +{\tau\over8}F_{ij}^2
           +{1\over 4\tau}(D_i\Phi)^2
        \right\} \,.
\end{equation}
The equations of motion corresponding to this Hamiltonian are
\begin{eqnarray}
 \dot{A_i} &=& {E_i\over \tau},\quad \dot{\Phi} = \tau p_{\eta},
  \quad (i=x,y),\nonumber\\
 \dot{E_x} &=& -\tau D_yF_{xy} + {ig\over\tau}[\Phi,D_x\Phi] ,\nonumber\\
 \dot{E_y} &=&  \tau D_xF_{xy} + {ig\over\tau}[\Phi,D_y\Phi] ,\nonumber\\
 \dot{p}_{\eta} &=& {1\over\tau}(D_x^2\Phi + D_y^2\Phi),\nonumber\\
\end{eqnarray}
and the Gauss law constraint is 
\begin{equation}
 \bm{D}\cdot \bm{E} +ig[p_{\eta},\Phi] = 0\, .
\end{equation}

\subsection{Discretized Hamiltonion formalism in 2+1 dimensions}

In order to realize numerically the solutions to the equations of motion 
in the previous section, while maintaining the gauge symmetry, we introduce the link variables
at the site $i$
\begin{equation}
    U_{j,i} = \exp\left[ i g a A_j(i) \right],\quad (j=x,y),
\end{equation}
where, $a$ is a lattice spacing.
Defining the plaquette
\begin{equation}
   U_{\Box} \equiv U_{l,j}U_{m,j+l}U^{\dagger}_{l,j+m}U^{\dagger}_{m,j}\,,
\end{equation}
the magnetic energy in the transverse plane is expressed for $a \to 0$ as
\begin{equation}
  {\tau \over 2a^2g^2}\sum_{\Box}(N_c - \rm{Re Tr} U_{\Box}),
\end{equation}
where the relation $U_{\Box} =\exp[ia^2gF_{\mu\nu}]$ was used.
The Hamiltonian on the lattice is 
\begin{eqnarray}
  H_L &=& 
    {1\over 4\tau}\sum_{\ell\equiv (j,\hat{n})}{\rm Tr}E_{\ell}^2
\nonumber \\ 
&+& {\tau\over 2}\sum_{\Box}(N_c-{\rm ReTr}U_{\Box}) \nonumber\\
&+& {1\over 4\tau}\sum_{j,\hat{n}}{\rm Tr}
           (\Phi_j - U_{\hat{n},j}
           \Phi_{j+\hat{n}} U^{\dagger}_{\hat{n},j})^2 \nonumber\\
&+& {\tau \over 4}\sum_{j}{\rm Tr}p_j^2 ,
\label{eq:HamiltonianDimLess}
\end{eqnarray}
where the following dimensionless variables are used
~\cite{Biro:1993qc,Heinz:1996wx}:
\begin{eqnarray}
   H & \to & {H\over g^2a},     \quad
   E \to {E\over ga^2},     \quad
   p \to {p\over ga},     \quad \\ \nonumber
 \Phi& \to& {\Phi\over ga},     \quad
  \tau \to a\tau .\quad 
\end{eqnarray}
Hamilton's equations of motion on the lattice are then 
\begin{eqnarray}
 \dot{U}_{r,i} &=& -{i\over\tau}U_{r,i}E_{r,i},
 \quad \dot{\Phi}_{i} = \tau p_{i},  \quad (r=x,y),\nonumber\\
 \dot{E}^a_{x,i} &=&  \tau {-i \over2} {\rm Tr} t^a\Big\{
   U_{y,i+\hat{x}}U^{\dagger}_{x,i+\hat{y}}
U^{\dagger}_{y,i}U_{x,i}\nonumber\\
 &-& U_{x,i}^{\dagger}U^{\dagger}_{y,i-\hat{y}}
     U_{x,i-\hat{y}} U_{y,i+\hat{x}-\hat{y}}\Big\}\nonumber\\
 &+& {ig\over\tau}[\Phi_{i+\hat{x}},U^{\dagger}_{x,i}\Phi_iU_{x,i}]
,\nonumber\\
 \dot{E}^a_{y,i} &=&  \tau {-i \over2} {\rm Tr} t^a\Big\{
  -U_{y,i+\hat{x}} U^{\dagger}_{x,i+\hat{y}}
U^{\dagger}_{y,i}U_{x,i}\nonumber\\
   &+& U_{x,i+\hat{y}-\hat{x}}^{\dagger}
       U^{\dagger}_{y,i-\hat{x}} U_{x,i-\hat{x}} U_{y,i} \Big\}\nonumber\\
    &+& {ig\over\tau}[\Phi_{i+\hat{y}},U_{y,i}^{\dagger}\Phi_iU_{y,i}] ,
               \nonumber\\
 \dot{p}_i &=& {1\over\tau}\Big[ \sum_{\hat{n}=x,y} \big(
          U_{\hat{n},i}\Phi_{\hat{n}+i} U_{\hat{n},i}^{\dagger} 
       +  U_{\hat{n},i-\hat{n}}^{\dagger}\Phi_{i-\hat{n}}
          U_{\hat{n},i-\hat{n}} \big)  \nonumber\\
  &-& 4\Phi_i
     \Big] .
\end{eqnarray}

Once one has explicit expressions for the fields and their canonically conjugate momenta,
other components of the Stress--Energy tensor can be computed as well.
In light-cone coordinates $(\tau, \eta,\vec{r})$,
the symmetric energy-momentum tensor can be defined as
\begin{equation}
 T^{\mu\nu} =  -{\tau\over2}\Tr(F^{\mu\alpha}F^{\nu}_{\ \alpha})
              - g^{\mu\nu} {\cal L}\, .
\end{equation}
The diagonal spatial components $T^{xx}$ and $T^{yy}$
 (the two components of the transverse pressure) are explicitly 
\begin{eqnarray}
 T^{xx} &=& {1\over4}\Tr \Big(
       {1\over\tau}(-E_x^2 + E_y^2  
   +  (D_x\Phi)^2 - (D_y\Phi)^2)\nonumber\\
    &+& \tau p_\eta^2
       + {\tau\over2}F_{ij}F_{ij}
                 \Big)  , \\
 T^{yy} &=& {1\over4}\Tr\Big(
      {1\over\tau}(+E_x^2 - E_y^2 
 - (D_x\Phi)^2 + (D_y\Phi)^2)\nonumber\\
      &+& \tau p_\eta^2
      + {\tau\over2}F_{ij}F_{ij}
                 \Big)  .
\end{eqnarray}
The $T^{\tau x}$ and $T^{\tau y}$ parts can be calculated as
\begin{equation}
 T^{\tau x} = -{1\over2}\Tr(
         E_{y}F_{xy} + p_{\eta}(D_x\Phi) ),
\end{equation}
$T^{\tau y}$ is obtained by replacing $x\leftrightarrow y$ in this expression.

The lattice equivalents of these continuum expressions are 
as follows. 
%
%
The transverse pressure in the $y$ direction is  
\begin{eqnarray}
  T^{yy}_i &=& 
    {1\over 8\tau}\sum_{j = i,i-\hat{x},i-\hat{y}}\Tr
        (E_{x,j}^2 - E_{y,j}^2)  \nonumber\\
 &+&  {\tau\over 8}\sum_{4\Box}(N_c-\ReTr U_{\Box}) \nonumber\\
 &-& {1\over 8\tau}\sum_{j=i,i-\hat{x}}\Tr
        (\Phi_j-U_{\hat{x},j}
            \Phi_{j+\hat{x}}U^{\dagger}_{\hat{x},j})^2 \nonumber\\
 &+& {1\over 8\tau}\sum_{j=i,i-\hat{y}}\Tr
        (\Phi_j-U_{\hat{y},j}
            \Phi_{j+\hat{y}}U^{\dagger}_{\hat{y},j})^2 \nonumber\\
 &+& {\tau \over 4}\Tr  p_i^2 \, .
\label{eq:tyyL}
\end{eqnarray}
where $4\Box$ means a sum over the four plaquettes which contain the link
$i$ with the same orientations.
$T^{xx}$ can be obtained by replacing $x\leftrightarrow y$
 in Eq.~(\ref{eq:tyyL}).

One of the ways to get the lattice definition of Poynting vector
may be a use of the conservation law $\partial_\mu T^{\mu\nu}=0$:
\begin{equation}
 \partial_{\tau} T^{\tau\tau}
           = \partial_{x}T^{\tau x}+\partial_{y}T^{\tau y},
\end{equation}
with the $T^{\tau\tau}$ which has to be defined in a local gauge invariant 
form at a site $i$ as
\begin{eqnarray}
  T^{\tau\tau}_i &=& 
    {1\over 8\tau}\sum^{'}_{\ell=(j,\hat{n})}{\rm Tr} E_{\ell}^2
 + {\tau\over 2}(N_c-{\rm ReTr}U_{\Box}) \nonumber\\
 & & + {1\over 4\tau}\sum^{'}_{j,\hat{n}}{\rm Tr}(\Phi_j-U_{\hat{n},j}
            \Phi_{j+\hat{n}}U^{\dagger}_{\hat{n},j})^2\nonumber\\
 & & + {\tau \over 16}\sum^{'}_{j}{\rm Tr}p_j^2 ,
\label{eq:t00area}
\end{eqnarray}
where $\sum^{'}_i \equiv i + (i+\hat{x}) + (i+\hat{y})$.
After quite some algebra, we obtain 
\begin{eqnarray}
T^{\tau x}_i &=& -{1\over4}E_{i,x}\Big[
                           U_{x,i+y}
                           U_{y,i+x}^{\dagger}
                           U_{x,i}^{\dagger}
                           U_{y,i} \nonumber\\
                       &+& ( U_{y,i}^{\dagger}
                             U_{x,i-x}^{\dagger}
                             U_{y,i-x}
                             U_{x,i-x+y} \Big]\nonumber\\
              &+& {1\over8}\Big[
               p_i( U_{x,i-x}^{\dagger}\Phi_{i-x}-U_{x,i}\Phi_i )\nonumber\\
              &+& p_{i+y}( U_{x,i+y-x}^{\dagger}\Phi_{i+y-x} \nonumber\\
              & & \qquad     - U_{x,i+y}\Phi_{i+y+x} ) \Big].
\end{eqnarray}


\begin{thebibliography}{99}

\bibitem{GLRMQ}
 L.V. Gribov, E. M. Levin and M. G. Ryskin, Phys. Repts. {\bf 100} (1983) 1;
 A. H. Mueller and J.-W. Qiu, Nucl. Phys. {\bf B268}(1986) 427;
 J. P. Blaizot and A. H. Mueller, Nucl. Phys. {\bf B289} (1987) 847.


\bibitem{MV}
L. McLerran and R. Venugopalan,
Phys. Rev. {\bf D49} 2233 (1994); {\bf D49} 3352 (1994); {\bf D50} 2225 (1994).

\bibitem{JKMW}
J. Jalilian--Marian, A. Kovner, L. McLerran and H. Weigert,  
  Phys. Rev. {\bf D55} 5414 (1997);
Y.~V.~Kovchegov, Phys.\ Rev.\ D {\bf 54}, 5463 (1996);  A.~Ayala, J.~Jalilian-Marian, L.~D.~McLerran and R.~Venugopalan,
Phys.\ Rev.\ D {\bf 52}, 2935 (1995); {\bf 53}, 458 (1996).

\bibitem{JIMWLK}
J. Jalilian-Marian, A. Kovner, A. Leonidov, and H. Weigert,
    Nucl. Phys. {\bf B504} 415 (1997);
J. Jalilian-Marian, A. Kovner, and H. Weigert,
    Phys. Rev. {\bf D59} 014015 (1999);
L. McLerran and R. Venugopalan,  Phys. Rev. {\bf D59} 094002 (1999);
E.~Iancu and L.~D.~McLerran,
Phys.\ Lett.\ B {\bf 510}, 145 (2001); E.~Iancu, A.~Leonidov and L.~D.~McLerran,
Nucl.\ Phys.\ A {\bf 692}, 583 (2001); E.~Iancu, A.~Leonidov and L.~D.~McLerran,
Phys.\ Lett.\ B {\bf 510}, 133 (2001).


\bibitem{CGC}L.~D.~McLerran,
arXiv:hep-ph/9903536;
Acta Phys.\ Polon.\ B {\bf 30}, 3707 (1999)
[arXiv:nucl-th/9911013];
arXiv:hep-ph/0104285;
E.~Iancu, A.~Leonidov and L.~McLerran,
arXiv:hep-ph/0202270.

\bibitem{RajGavai}
R. V. Gavai and R. Venugopalan, Phys.\ Rev.\ {\bf D54}, 5795 (1996).

\bibitem{Mueller1}
A. H. Mueller, Nucl. Phys. {\bf B572} (2000) 227.


\bibitem{ActPol}
R.~Venugopalan,
Acta Phys.\ Polon.\ B {\bf 30}, 3731 (1999).

\bibitem{KN}D. Kharzeev and M. Nardi, Phys.\ Lett.\ B {\bf 507}, 121 (2001).


\bibitem{AYR02}
A.~Krasnitz, Y.~Nara and R.~Venugopalan,
arXiv:hep-ph/0204361; D.~Teaney and R.~Venugopalan,
Phys.\ Lett.\ B {\bf 539}, 53 (2002).

\bibitem{KMW}
A. Kovner, L. McLerran and H. Weigert,
 Phys. Rev {\bf D52} 3809 (1995); {\bf D52} 6231 (1995).

\bibitem{DYEA}
 Y. V. Kovchegov and D. H. Rischke,  Phys. Rev.  {\bf C56} (1997) 1084;
 S. G. Matinyan, B. M\"uller and D. H. Rischke,
 Phys. Rev. {\bf C56} (1997) 2191;  Phys. Rev. {\bf C57} (1998) 1927;
 Xiao-feng Guo,  Phys. Rev. {\bf D59} 094017 (1999).

\bibitem{GyulassyMcLerran}
 M. Gyulassy and L. McLerran, Phys. Rev.  {\bf C56} (1997) 2219.

\bibitem{Balitsky}I.~Balitsky,
arXiv:hep-ph/0101042.


\bibitem{AR99}
A. Krasnitz and R. Venugopalan, hep-ph/9706329, hep-ph/9808332;
Nucl. Phys. {\bf B557} 237 (1999).

\bibitem{AR00}
A. Krasnitz and R. Venugopalan, Phys. Rev. Lett. {\bf 84} (2000) 4309.

\bibitem{AR01}
A. Krasnitz and R. Venugopalan, Phys. Rev. Lett. {\bf 86} (2001) 1717.

\bibitem{AYR01}
A. Krasnitz, Y. Nara and R. Venugopalan,
 Phys.\ Rev.\ Lett.\ {\bf 87}, 192302 (2001).  


\bibitem{Kogut}
 J. Kogut and L. Susskind, Phys. Rev. {\bf D11}, 395 (1975);
 J. B. Kogut, Rev. Mod. Phys. {\bf 51}, 659 (1979);
 J. B. Kogut, Rev. Mod. Phys. {\bf 55}, 775 (1983).


\bibitem{BMSS}
R. Baier, A. H. Mueller, D. Schiff and D. T. Son,
 Phys. Lett. {\bf B502} 51 (2001).

\bibitem{BMSS2}R.~Baier, A.~H.~Mueller, D.~Schiff and D.~T.~Son,
Phys.\ Lett.\ B {\bf 539}, 46 (2002).

\bibitem{MBVDG}
A. H. Mueller, Phys. Lett. {\bf B475} 220 (2000);  
J. Bjoraker and R. Venugopalan,  Phys. Rev. {\bf C63} 024609 (2001);  
A. Dumitru and M. Gyulassy, Phys. Lett. {\bf B494}, 215 (2000).

\bibitem{Bass}
S.~A.~Bass, B.~Muller and D.~K.~Srivastava,
arXiv:nucl-th/0207042; D.~Molnar and M.~Gyulassy,
Phys.\ Rev.\ C {\bf 62}, 054907 (2000).

\bibitem{Nara:2001zu}
Y.~Nara, S.~E.~Vance and P.~Csizmadia,
Phys.\ Lett.\ B {\bf 531}, 209 (2002)
[arXiv:nucl-th/0109018].

\bibitem{ShiMuller}
G.~R.~Shin and B.~Muller,
arXiv:nucl-th/0207041.

\bibitem{KNL}D. Kharzeev and E. Levin, Phys.\ Lett.\ B {\bf 523}, 79 (2001); 
D.~Kharzeev, E.~Levin and M.~Nardi,
arXiv:hep-ph/0111315.

\bibitem{Juergen}J.~Schaffner-Bielich, D.~Kharzeev, L.~D.~McLerran and R.~Venugopalan,
Nucl.\ Phys.\ A {\bf 705}, 494 (2002).



\bibitem{DJLT}A.~Dumitru and J.~Jalilian-Marian,
Phys.\ Rev.\ Lett.\  {\bf 89}, 022301 (2002); F.~Gelis and J.~Jalilian-Marian,
Phys.\ Rev.\ D {\bf 66}, 014021 (2002); J.~T.~Lenaghan and K.~Tuominen,
arXiv:hep-ph/0208007.


\bibitem{KovMuell}
Y.~V.~Kovchegov and A.~H.~Mueller,
Nucl.\ Phys.\ B {\bf 529}, 451 (1998).

\bibitem{Dumitru}
A.~Dumitru and L.~D.~McLerran,
Nucl.\ Phys.\ A {\bf 700}, 492 (2002).


\bibitem{LM}
C.~S.~Lam and G.~Mahlon,
Phys.\ Rev.\ D {\bf 61}, 014005 (2000); Phys.\ Rev.\ D {\bf 62}, 114023 (2000); 
Phys.\ Rev.\ D {\bf 64}, 016004 (2001).






\bibitem{ILM}E.~Iancu and L.~D.~McLerran,
Phys.\ Lett.\ B {\bf 510}, 145 (2001).

\bibitem{Mueller2002}A.~H.~Mueller,
arXiv:hep-ph/0206216.



\bibitem{phobos130}
B.~B.~Back {\it et al.}  [PHOBOS Collaboration],
arXiv:nucl-ex/0105011.

\bibitem{phenix_mult}
K.~Adcox {\it et al.}  [PHENIX Collaboration],
Phys.\ Rev.\ Lett.\  {\bf 86}, 3500 (2001)
[arXiv:nucl-ex/0012008].
%

\bibitem{phenix_et}
K.~Adcox {\it et al.}  [PHENIX Collaboration],
Phys.\ Rev.\ Lett.\  {\bf 87}, 052301 (2001)
[arXiv:nucl-ex/0104015].
%


\bibitem{KIIM}E.~Iancu, K.~Itakura and L.~McLerran,
Nucl.\ Phys.\ A {\bf 708}, 327 (2002).







\bibitem{cteq5}
H.~L.~Lai {\it et al.}  [CTEQ Collaboration],
Eur.\ Phys.\ J.\ C {\bf 12}, 375 (2000) 
[arXiv:hep-ph/9903282].


\bibitem{pythia}
T.~Sj\"ostrand, P.~Ed\'en, C.~Friberg, L.~L\"onnblad, G.~Miu, S.~Mrenna
and E.~Norrbin, Comp. Phy. Commun. {\bf 135} (2001) 238.

\bibitem{ua1}
C.~Albajar {\it et al.}  [UA1 Collaboration],
Nucl.\ Phys.\ B {\bf 335}, 261 (1990).


\bibitem{Ad:note}see Eq.~(73) of Ref.~\cite{Dumitru}.

\bibitem{comment1}One must caution that the considerations in this section assume that both $\Lambda_{s1}$ 
and $\Lambda_{s2}$ are large and much greater than $\Lambda_{QCD}$. For these asymptotics to apply at current 
energies, one would have to assume a range of validity for the model beyond what our naive estimates would 
suggest.


\bibitem{AYR03}A. Krasnitz, Y. Nara and R. Venugopalan, in progress.

\bibitem{KT}Y.~V.~Kovchegov and K.~L.~Tuchin,
Nucl.\ Phys.\ A {\bf 708}, 413 (2002).








\bibitem{Biro:1993qc}
T.~S.~Biro, C.~Gong, B.~Muller and A.~Trayanov,
Int.\ J.\ Mod.\ Phys.\ C {\bf 5}, 113 (1994).

\bibitem{Heinz:1996wx}
U.~W.~Heinz, C.~R.~Hu, S.~Leupold, S.~G.~Matinian and B.~Muller,
Phys.\ Rev.\ D {\bf 55}, 2464 (1997).

\end{thebibliography}
\end{document}